\documentclass[12pt,preprint]{aastex}
\slugcomment{Draft writing}
\shorttitle{FIREBALL GENERATION}
\shortauthors{ASANO and TAKAHARA}

\begin{document}

\title{GENERATION OF A FIREBALL IN AGN HOT PLASMAS}
\author{\scshape K. Asano}
\affil{Division of Theoretical Astronomy, National Astronomical Observatory of Japan}
\email{asano@th.nao.ac.jp}
\and
\author{\scshape F. Takahara}
\affil{Department of Earth and Space Science, Graduate School of Science, Osaka University}
\email{takahara@vega.ess.sci.osaka-u.ac.jp}

\date{Submitted; accepted }

\begin{abstract}

Motivated by relativistic jets observed in active galactic nuclei (AGN),
we simulate outflows of electron-positron pairs strongly coupled
with photons from normal electron-proton plasmas.
Using multi-fluid approximation and a Monte Carlo method of radiative 
transfer, we obtain spherically symmetric,
steady solutions of radiation and pair outflows
for the luminosity $L \leq 10^{47}$ erg ${\rm s^{-1}}$.
For microphysics, Coulomb scattering, Compton scattering,
bremsstrahlung, electron-positron
pair annihilation and creation are taken into account.
Although a significant amount of pairs outflow
by powerful radiative force with a mildly relativistic velocity,
the temperature is not high enough to avoid pair annihilation 
before the fireball becomes optically thin to scattering. 
Several caveats in the simulations are also discussed. 
\end{abstract}


\section{INTRODUCTION}
\label{INTRODUCTION}

Relativistic jets are observed in active galactic nuclei (AGNs) and
Galactic black hole candidates.
The velocity of these jets is highly relativistic
with a bulk Lorentz factor above 10 and 
the kinetic power is almost comparable to the
Eddington luminosity.
The production and bulk acceleration mechanisms of these jets
are still unknown, though many ideas have been
proposed ranging from magneto-hydrodynamical to
radiative ones.

Although it is difficult to determine the matter content of jets
from observations, several independent arguments favor electron-positron jets
\citep{tak94,tak97,rey96,war98,hom99,hir99,hir00,hir05,kin04,cro05}.
Electron-positron jets are most likely produced
in accretion disks around the central black holes.
Because the electron mass is much smaller than the proton mass,
the produced electron-positron pairs can be more easily ejected 
than protons.
Some papers discuss the accretion disks
with electron-positron outflows \citep{mis95,lia95,li96,yam99}.
If the accretion disks form hot pair plasma
strongly coupled with photons,
the plasma may be thermally accelerated like the fireball
applied to gamma-ray bursts (GRBs) \citep{ree92}.
As long as the initial conditions for the fireball model
are satisfied, the flow accelerates to a relativistic velocity
undoubtedly \citep{pir93,asa02}.
However, the characteristic size of AGNs is too large
to form fireballs which are in a complete thermal equilibrium
at high temperatures comparable to the electron rest mass energy
$m_{\rm e} c^2$.
In the standard fireball model, to make matters worse,
electron-positron pairs are almost wholly annihilated
in the course of the thermal expansion.

To overcome this situation, \citet{iwa02,iwa04} showed 
that a ``Wien fireball'',
which is optically thick to Compton scattering but thin
to absorption, results in a relativistic outflow
avoiding the difficulties of pair annihilation.
When high-energy photons are provided sufficiently,
copious electron-positron pairs are produced.
From the size and luminosity of AGNs, the pair plasma
is not expected to be in a complete thermal equilibrium.
The pair plasma may be optically thin to absorption but
thick to scattering.
Since photons and pairs are coupled by lepton scattering,
the plasma can be thermally accelerated expending its internal energy.
In this case, as long as the temperature remains relativistic
($\gtrsim m_{\rm e} c^2$),
sufficient pair creation by high-energy photons
compensates for pair annihilation.
Since the cross section of pair annihilation and that of
Compton scattering are the same order,
the number of pairs is almost conserved
outside the photosphere, where
photons and pairs are decoupled.
Therefore, if the temperature at the photosphere
is relativistic, a sufficient amount of electron-positron pairs
survive as a relativistic outflow.
Although super-Eddington luminosity in spherical symmetry
is needed in the Wien fireball model,
the real luminosity for collimated 
jets can be sub-Eddington.

Pairs are formed via photon photon collisions
in the accretion disk composed of normal plasma (electron-proton).
Assuming that pairs are confined with protons,
several authors \citep{lig82,sve82,sve84,zdz84,kus85,kus87}
investigated pair equilibrium plasmas.
As is typically shown in \citet{kus87},
the plasma temperature becomes $\sim 0.1 m_{\rm e} c^2$
for plasmas of size $\sim 10^{14}$ cm
and luminosity $\gtrsim 10^{45} {\rm ergs}$ ${\rm s}^{-1}$,
because of the high cooling rate of dense pairs.
On the other hand, the Wien fireball model
requires a super Eddington luminosity
$\gtrsim 10^{47}$ erg ${\rm s^{-1}}$ in the spherical symmetry
and a high temperature $>0.5 m_{\rm e} c^2$.
The series of studies of pair equilibrium plasmas
imply that the equilibrium temperature is too low 
for the Wien fireball model.

However, considering the effects of pair escape from the disk
by their own pressure or radiative force \citep{yam99},
there is a possibility that
the density and temperature of pairs are drastically different
from the values obtained assuming static plasmas.
In order to prove pair ejection from the disk,
one may need to treat multi-plasma dynamics with radiation field.
However, there has been no study of formation and ejection of
pairs from the accretion disk taking into account radiative
transfer consistently.

In this paper we simulate outflows of electron-positron pairs
from hot plasmas with radiative transfer in order to
discuss whether or not AGNs can produce fireballs with
sufficiently high temperatures.
For simplicity, the multi-fluid approximation is adopted, and
the plasma is assumed to be spherically symmetric.
The microscopic physical processes we take into account
are Coulomb scattering, Compton scattering,
bremsstrahlung, electron-positron
pair annihilation and creation in electron, positron,
and proton plasmas.
In \S 2 we explain our method, and the numerical results are
shown in \S 3.
\S 4 is devoted to summary and discussion.

\section{METHOD}

In this paper we numerically obtain spherically symmetric,
steady solutions of radiation and pair outflows from hot plasmas 
undergoing steady external heating.
First of all we consider a static
plasma consisting of protons ({\it p}) and electrons ({\it e}).
The proton number density is assumed to be
\begin{equation}
n_{\it p}(R) =
n_0 \exp(-(R/R_0)^2),
\end{equation}
where $R$ is the radius from the center.
The total heating rate of the plasma is given as a parameter $L$.
Assuming that the heating rate is proportional to $n_{\it p}$,
the proton heating rate per unit volume is obtained as
\begin{equation}
H_0(R) =
\frac{L}{\pi^{3/2} R_0^3} \exp(-(R/R_0)^2).
\end{equation}
Inside this plasma electron-positron pairs ({\it e$^\pm$}) are
produced via photon photon collisions.
In this calculation we assume the plasmas are divided into three fluids:
proton ({\it p}), background electron ({\it e}), and pair ({\it e$^\pm$}) fluids.
The background electrons are confined with protons to maintain charge neutrality.
Thus, the number density of the {\it e}-fluid, $n_{\it e}$, is
the same as that of the {\it p}-fluid, $n_{\it p}$.
On the other hand, positrons
may move independently of the {\it p}-fluid.
In order to maintain charge neutrality
positrons of the density $n_+$ in the comoving frame
accompany electrons of the same density $n_-=n_+$.
Namely, electrons are artificially divided into two components;
one belongs to the {\it e}-fluid, and another belongs to the pair-fluid.
Even if the pair-fluid flows with velocity $\beta \equiv v/c$,
{\it p} and {\it e}-fluids are assumed to be static.
Although the validity of this multi-fluid approximation is not always assured,
we adopt this approximation for simplicity.

The conservation laws of energy, momentum, and number of electron-positron
pairs
in the rest frame of {\it p}-fluid $K$ are given by
\begin{eqnarray}
\frac{1}{R^2} \frac{d}{dR} \left[
R^2 (\epsilon_\pm+P_\pm) \Gamma U
\right]=H,
\label{H}
\end{eqnarray}
\begin{eqnarray}
\frac{1}{R^2} \frac{d}{dR} \left[
R^2 (\epsilon_\pm+P_\pm) U^2
\right]+\frac{d P_\pm}{dR}=F,
\end{eqnarray}
\begin{eqnarray}
\frac{1}{R^2} \frac{d}{dR} \left[ R^2 n_\pm U \right]=Q,
\label{Q}
\end{eqnarray}
where $n_\pm=2 n_+$, $\epsilon_\pm$ and $P_\pm$ are the number density, 
energy density and pressure of
the pair-fluid in the fluid rest frame $K'$, respectively.
The velocity of the pair-fluid is
represented by the Lorentz factor, $\Gamma=1/\sqrt{1-\beta^2}$,
and 4-velocity $U=\sqrt{\Gamma^2-1}$.
The pressure is written as $n_\pm T_\pm$,
where $T_\pm$ is the temperature of the pair-fluid.
Normalizing $T_\pm$ by the electron mass as $\theta_\pm=T_\pm
/(m_{\it e} c^2)$,
the energy density in the rest frame,
$\epsilon_\pm$,
is written as
$n_\pm m_{\it e} c^2 (K_3(1/\theta_\pm)/K_2(1/\theta_\pm)
-\theta_\pm)$,
where $K_i$ is the $i$th order modified Bessel function of the second kind.
The source terms $H$, $F$, and $Q$,
which are measured in the coordinate frame,
are explained in the following subsections.

\subsection{Energy Budget}

The energy source term for pair-fluid $H$ is expressed as
$H=H_{\it p}+H_{\it e}+H_\gamma$,
where $H_{\it p}$ and $H_{\it e}$ are the heating rates by protons
and background electrons via Coulomb scattering, respectively,
$H_\gamma$ is the heating rate due to interactions with photons.

When there exists relative velocity
between the pair-fluid and background fluids,
it may be hard to obtain $H_{\it p}$ and $H_{\it e}$ analytically.
\citet{asa06} numerically calculated $H_{\it p}$ and $H_{\it e}$
for two Maxwell-Boltzmann gasses
with relative velocities $U=10^{-2}$-$10^2$
using relativistic formulae.
Since these rates are functions of $U$, $T_\pm$,
and background proton (electron) temperature $T_{\it p}$
($T_{\it e}$), we have no simple fitting formulae
for $H_{\it p}$ and $H_{\it e}$.
However, if $U \lesssim 1$, numerically obtained $H_{\it p}$
agrees rather well with the analytical formula of $U=0$,
\begin{eqnarray}
H_{\it p}=n_\pm n_{\it p} \frac{3 m_{\it e} \sigma_{\rm T}}{2 m_{\it p}}
\frac{T_{\it p}-T_\pm}
{K_2(1/\theta_\pm) K_2(1/\theta_{\it p})} \ln{\Lambda} \times \nonumber \\
\left[
\frac{2 (\theta_\pm+\theta_{\it p})^2+1}{\theta_\pm+\theta_{\it p}}
K_1 \left(\frac{\theta_\pm+\theta_{\it p}}{\theta_\pm \theta_{\it p}}\right)
+2 K_0 \left( \frac{\theta_\pm+\theta_{\it p}}{\theta_\pm \theta_{\it p}}\right)
\right],
\label{Hp}
\end{eqnarray}
\citep{ste83b},
where $\theta_{\it p} \equiv T_{\it p}/(m_{\it p} c^2)$
and $\ln{\Lambda}$ is the Coulomb logarithm (we adopt $\ln{\Lambda}=20$ hereafter). 
In our simulation we use the above formula for $H_{\it p}$
and tables of the case of $U=0$ in \citet{asa06} for $H_{\it e}$, irrespective of $U$.
As will be shown later in our results,
$U$ becomes $\sim 1$ in low-density regions, where the Coulomb energy transfer
is not effective.
Therefore, we neglect the effect of finite $U$ on $H_{\it p}$
and $H_{\it e}$ in our case.

The heating rate of pair-fluid through photons $H_\gamma$ is divided 
into two parts
as $H_\gamma=H_{\rm cir}-H_{\rm rad}$,
where $H_{\rm rad}$ is the cooling rate due to
photon production (bremsstrahlung and pair annihilation), and
$H_{\rm cir}$ is the heating rate by circumambient photons
via Compton scattering and pair-creation.
In this paper we only consider thermal bremsstrahlung and pair-annihilation
for photon-production processes.
We use the formulae of the cooling rates in \citet{sve82}
for bremsstrahlung ({\it ep}, {\it ee}, and {\it e$^+$e$^-$})
and pair annihilation.
The radiative cooling rates due to scattering of two particles
in the background fluids ({\it ep} and {\it ee}-bremsstrahlung)
are easily obtained.
Hereafter, we call such emissions ``background-fluids emissions''.
Given the temperature $T_{\it e}$, these cooling rates via bremsstrahlung,
$B^{\rm bg}_{\it ep}$ and $B^{\rm bg}_{\it ee}$
$[{\rm erg}~{\rm cm}^{-3}~{\rm s}^{-1}]$,
are proportional to $n_{\it p}^2$,
where the superscript ``bg'' designates the background fluids.

On the other hand, bremsstrahlung emission and pair annihilation
of two particles in the pair-fluid
(pair-fluid emissions) are calculated in the comoving frame $K'$.
The Lorentz transformation implies that
the cooling rates due to bremsstrahlung, $B^{\rm pair}_{\it +-}$,
and $B^{\rm pair}_{\it --}$=$B^{\rm pair}_{\it ++}$,
and the rate due to pair annihilation, $A^{\rm pair}$
in the coordinate frame $K$ are $\Gamma$ times those in the frame $K'$.

We should notice that there are emission processes
between two particles of different fluids;
one belongs to the pair-fluid and
another belongs to the background fluids.
We call such emissions ``inter-fluid emissions''.
Since the pair-fluid is moving velocity,
inter-fluid emissions are not isotropic in any frame.
In our simulation we adopt the following approximation
for the cooling rates due to inter-fluid emissions.
In the frame $K_*$, which moves with velocity $\beta_*=U/(1+\Gamma)$,
the two fluids flow symmetrically;
the background fluids flow with velocity $\beta_*$,
while the pair-fluid flows towards the opposite direction
with the same speed $\beta_*$.
In this frame the densities of the {\it e}-fluid
and pair-fluid are $n_{\it p*}=\Gamma_* n_{\it p}$
and $n_{\pm *}=\Gamma_* n_\pm$, respectively, where $\Gamma_*=1/\sqrt{1-\beta_*^2}$.
We use the usual formulae for thermal and isotropic plasma
adopting the temperature as $\max{(T_{\rm av},\beta_* \Gamma_*)}$
and densities $n_{\it p*}$ and $n_{\pm *}$, where
$T_{\rm av} \equiv \Gamma_* (T_\pm+T_{\it e})/2$
for bremsstrahlung, $B^{\rm int}_{e-}$ and $B^{\rm int}_{e+}$,
and pair annihilation, $A^{\rm int}$,
or $T_{\rm av} \equiv \Gamma_* T_\pm$ for
bremsstrahlungs, $B^{\rm int}_{p-}=B^{\rm int}_{p+}$.
The superscript ``int'' means inter-fluid emissions.
The cooling rates in the frame $K$ become
$\Gamma_*$ times those in the frame $K_*$.
We assume that two fluids lose energies equally
via these emissions, except for $B^{\rm int}_{p-}$
and $B^{\rm int}_{p+}$ processes, in which only the pair-fluid
loses energy.

The above approximation may seem rough.
However, as will be shown in our results,
$n_\pm \gg n_{\it p}$ and $T_\pm \sim T_{\it e}$
are plausible for energetic jets.
Therefore, inter-fluid emissions
have minor effects on the motion of the pair-fluid.

From the above assumptions we obtain
\begin{eqnarray}
H_{\it rad}=\frac{1}{c} \left( A^{\rm pair}+\frac{1}{2} A^{\rm int}
+2 B^{\rm int}_{p-}+2 B^{\rm pair}_{\it ++}+ \right. \nonumber \\
\left. \frac{1}{2} B^{\rm int}_{e-}
+ \frac{1}{2} B^{\rm int}_{e+} + B^{\rm pair}_{+-} \right).
\end{eqnarray}

The heating rate $H_{\rm cir}$ by circumambient photons
via pair creation and Compton scattering
is obtained from calculation of radiative transfer,
which is explained in the next subsection.

The heating rate of background electrons, $H_{\it ep}$,
by protons is also obtained from (\ref{Hp}) replacing $T_\pm$ and $n_\pm$
by $T_{\it e}$ and $n_{\it p}$, respectively.
Then, we can obtain the temperature of protons $T_{\it p}$
by energy balance equation as
\begin{eqnarray}
H_0=H_{\it p}+H_{\it ep}.
\label{H0}
\end{eqnarray}
The background electrons are also heated by photons
via Compton scattering.
We write the heating rate of {\it e}-fluid by circumambient photons
as $H_{{\it e}, {\rm cir}}$.
Then, the temperature of {\it e}-fluid, $T_{\it e}$, is obtained from
\begin{eqnarray}
H_{\it ep}-H_{\it e}+H_{{\it e}, {\rm cir}}=\frac{1}{c} \left( \frac{1}{2} A^{\rm int}
+B^{\rm bg}_{\it ep}+B^{\rm bg}_{\it ee}+ \right. \nonumber \\
\left. \frac{1}{2} B^{\rm int}_{e-}
+ \frac{1}{2} B^{\rm int}_{e+} \right).
\label{He}
\end{eqnarray}

\subsection{Radiative Transfer}

We solve radiative transfer with the Monte Carlo method,
taking into account pair-creation and Compton scattering.
Our numerical method is basically the same as \citet{iwa04}.
We shortly explain the method in this subsection.

We divide the calculation regime into many spherical shell elements.
From each shell, photons are emitted via
background, pair, and inter-fluid emissions,
which are assumed to be isotropic in the frames,
$K$, $K'$, and $K_*$, respectively.
For bremsstrahlung, we adopt an approximation
of number spectra $\propto \varepsilon^{-1.12} \exp{[-(2 \varepsilon/T)^{0.8}]}$,
where $T=T_{\it e}$, $T_\pm$, or $T_{\rm av}$ for each emission process,
normalized by the emissivities discussed in the former subsection.
This approximation well reproduces the specral shapes of
bremsstrahlung numerically obtained, based on the methods of quantum
electrodynamics \citep{hau75,hau85,hau03}.
On the other hand, we adopt a fitting formula of \citet{zdz80}
for pair-annihilation spectrum.

The pair-fluid emission and the inter-fluid emission
are assumed to be isotropic in the frames $K'$ and $K_*$, respectively.
We define the emission coefficient, $j_\varepsilon(\mu)$ [${\rm cm}^{-3}$
${\rm s}^{-1}$ ${\rm sr}^{-1}$]: emitted energy per unit time
per unit volume per unit solid angle and per unit photon-energy,
where $\mu$ is the cosine between the direction of the photon
and the radial direction.
Taking the inter-fluid emissions as an example,
the emission coefficient in the frame $K$ is obtained from
the Lorentz invariant $j_\varepsilon/\varepsilon^2=j_{\varepsilon *}/
\varepsilon_*^2$ \citep{ryb79} and $\varepsilon_*=\varepsilon \Gamma_*
(1-\beta_* \mu)$, where $\varepsilon_*$ is the photon-energy
in the frame $K_*$.
For the pair-fluid emissions too, we obtain the emission coefficients
in the same way.

Generating photons according to the above method,
we simulate the trajectory of each photon propagating
through the three fluids and photon field,
where the photon interacts with electrons, positrons,
and photons themselves.
Elementary processes to be considered are Compton scattering
and pair creation.
We use the Klein-Nishina formula and the representation in \citet{gou67}
for the cross sections of Compton scattering and pair creation, respectively.
We simulate all these processes by the Monte Carlo method.
Details of the method of following the photon trajectory is given in \citet{iwa04}.

After simulating all photons emitted from all shells toward all directions,
we numerically obtain the photon distribution
in the phase space, the heating rates $H_{\rm cir}$ and
$H_{{\it e}, {\rm cir}}$,
the radiative force on the pair-fluid $F_\gamma$, and the pair creation rate
$Q_\gamma$.
However, results depend on the photon distribution assumed in advance.
Therefore, given the parameters of the three fluids,
we simulate iteratively until the photon distribution converges.

\subsection{Momentum and Number Sources}

The momentum source $F$ is expressed as $F=F_\gamma-F_{\rm rea}-F_{\it e}-F_{\it p}$,
where $F_\gamma$ is the radiative force due to pair creation and Compton scattering,
$F_{\rm rea}$ is the reaction force due to photon-emission,
and $F_{\it e}$ and $F_{\it p}$ are the frictional forces due to Coulomb scattering
with background electrons and protons, respectively.
As was mentioned in the former subsection,
$F_\gamma$ is obtained from the calculation of radiative transfer.

\citet{asa06} calculated $F_{\it e}$ and $F_{\it p}$ numerically
for two Maxwell-Boltzmann gasses
with relative velocities $U=10^{-2}$-$10^2$
using relativistic formulae,
and they express their results as a form
\begin{equation}
F_{\it e,p}=(\Gamma n_\pm) n_{\it p} m_{\rm e} c^2 f_{\it e,p}(T_\pm,
T_{\it e,p},U),
\end{equation}
where $\Gamma n_\pm$ is the density of pairs in the frame $K$.
The functions $f_{\it e,p}$ are listed as tables in \citet{asa06}.
For $U \ll 1$, $f_{\it e,p}$ increases with $U$ monotonically.
Beyond $U \sim 1$, $f_{\it p}$ decreases with $U$,
while $f_{\it e}$ is nearly constant.
The lower temperatures $T_\pm$ and $T_{\rm e}$ become,
the larger the frictional forces become.

Although photons are emitted isotropically in the frame $K'$,
the photons are beamed in the frame $K$ and the pair-fluid
loses momentum.
The Lorentz transformation implies
\begin{eqnarray}
F_{\rm rea}=\frac{\beta}{c} \left( A^{\rm pair}+2 B^{\rm pair}_{\it ++}
+ B^{\rm pair}_{+-} \right),
\label{frad}
\end{eqnarray}
where we neglect the contribution of the inter-fluid emission.
In the strict sense the inter-fluid emission causes momentum transfer
between the pair-fluid and the background fluids.
In order to obtain the momentum transfer in
bremsstrahlung, the scattering processes of
two particles with photon emission, we need messy and heavy
calculations using the differential cross sections as
functions of both the particle scattering angle and the photon angle.
On the other hand, the momentum transfers due to pair-annihilation
are numerically calculated in \citet{asa06}, which shows
that the rate due to pair-annihilation is negligible
in comparison with that due to Coulomb scattering.
The momentum transfers due to processes accompanying photon emission,
including bremsstrahlung, may be negligible.
The smallness of momentum transfer
and the difficulties in calculations are reasons why we have neglected
the contribution of the inter-fluid emission in equation (\ref{frad}).

Of course, the background fluids are also pushed outward
by photons and the frictional force.
In this paper, we assume that the background fluids are maintained 
to be static; 
such forces are balanced with the gravitational
force by the central black hole, which is implicit in 
our geometry of computation.
The {\it e}-fluid suffers from
the counteractive frictional force $F_{\it e}$ and the radiative force,
which can be stronger than the total force on the {\it p}-fluid.
There should be a certain degree of
charge separation between the {\it e}-fluid and {\it p}-fluid
to maintain the {\it e}-fluid static.
The resultant static electric field may affect the motion
of the pair-fluid.
Electrons in the pair-fluid are decelerated, while positrons
are accelerated.
The force due to the static electric field on the pair-fluid
becomes zero on the average.
As long as we adopt the multi-fluid approximation,
we can neglect the effect of the static electric field,
although there is possibility that the kinematical effects of the
electric field heat the pair-fluid.

The source term $Q$, which is Lorentz invariant,
is expressed as $Q=Q_\gamma-Q_{\rm ann}$,
where $Q_\gamma$ and $Q_{\rm ann}$ are the pair creation and annihilation rates,
respectively.
While $Q_\gamma$ is obtained from the calculation of radiative transfer,
we adopt a formula in \citet{sve82} for $Q_{\rm ann}$.
As is the case of radiative cooling, the positron annihilation rate
is divided into two parts as $Q_{\rm ann}=Q_{\rm ann}^{\rm pair}+Q_{\rm ann}^{\rm int}$.
The method of calculation of $Q_{\rm ann}^{\rm int}$ is the same way
as the inter-fluid cooling, in which $n_{\it p *}$, $n_{\pm *}$, and $T_{\rm av}$
are used.
Even if a positron annihilates with an electron in the {\it e}-fluid,
the number density of {\it e}-fluid should be invariant
to maintain the multi-fluid approximation.
Therefore, we should consider that an electron in the pair-fluid
is traded to the {\it e}-fluid in this case.
As a result two particles in the pair fluid disappear
in each inter-fluid annihilation.
The trade of electrons means that energy and momentum
are also transferred between the two fluids.
However, as we have mentioned several times,
such energy and momentum transfer may be negligible in this case too.
Although this unavoidable treatment in the inter-fluid annihilation
indicates a limit in the multi-fluid approximation,
we adopt this approximation to save cost in computation.

\subsection{Flow Solution}

We solve $n_\pm$, $\Gamma$, $T_\pm$, $T_{\it e}$, and $T_{\it p}$
from equations (\ref{H})-(\ref{Q}),(\ref{H0}), and (\ref{He}).
First of all we set plausible values for the three fluids.
Then, combined with radiative transfer to obtain the source terms,
we mimic the time evolution of the plasmas.
Since the photon diffusion time scale is much longer than
the dynamical time scale, we cannot follow the time evolution
of both the plasmas and photon field consistently.
However, the final steady solutions are consistently obtained
by mimicking the time evolution for a sufficiently long time.

\section{NUMERICAL RESULTS}

In our numerical simulation,
the energy space of photons $x \equiv \varepsilon/m_{\rm e} c^2$
is divided into 70 bins in the logarithmic manner between
$10^{-5}$ and $10^2$.
For the direction of photons, $\mu$ is divided into 20 bins
between $-1$ and 1.
The outer boundary is set at $R=R_{\rm out} \equiv 2 R_0$,
and the radial coordinate is divided into 40 shells.
We follow more than $10^5$ paths of photons to solve
radiative transfer.
Some figures in this section are plotted
using results after following $5.6 \times 10^7$ paths of photons.
We do not take into account photons from outside
of the boundary.
However, such photons may not affect the outflow as will be shown later.

In this calculation the parameters are the radius $R_0$,
density $n_0$ and luminosity $L$.
For several values of $L$ above $10^{44}$ erg ${\rm s^{-1}}$,
we simulate plasmas for two sets of parameters: $R_0=10^{14}$ cm with
$n_0=10^{10}$ ${\rm cm^{-3}}$,
and $R_0=3 \times 10^{14}$ cm with $n_0=\frac{1}{3} \times 10^{10}$
${\rm cm^{-3}}$.
These values are characteristic of accretion plasmas
in AGNs.
In both  cases, the pair free Thomson scattering depth
$\tau_{\rm p} \sim n_0 R_0 \sigma_{\rm T} \sim 0.7$.

\begin{figure}[t]
\centering
\epsscale{1.0}
\plotone{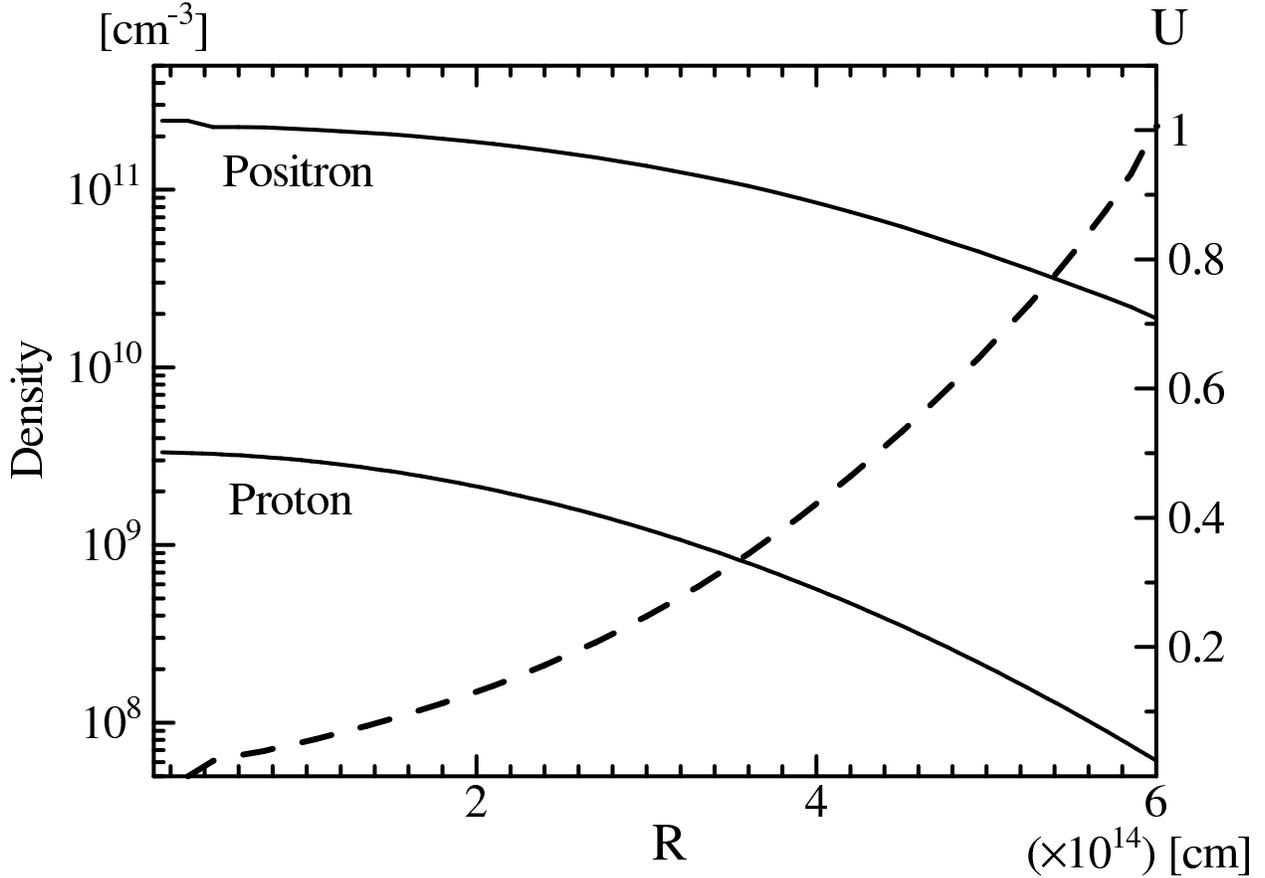}
\caption{
Positron density $n_+$ (solid, left axis) in the frame $K'$
and 4-velocity $U$ (dashed, right axis).
The proton density is also plotted for reference.
}
\end{figure}

When $L$ is less than $\sim 10^{45}$ erg ${\rm s^{-1}}$,
the results are almost the same as those in \citet{kus87}.
As the luminosity $L$ increases,
the temperature of pairs decreases,
while the pair density increases.
The positron density is comparable to the proton density.
However, a depression of the pair density in the central region,
which is seen in \citet{kus87} for $L \sim 10^{45}$ erg ${\rm s^{-1}}$,
is not observed in our simulation.
The pair density decreases with radius $R$ monotonically.
The pair fluid escapes from the boundary by their own pressure.
The radiative force is not effective to drive the outflow in these cases.
However, the energy fraction of the outflow is negligible
compared with the total luminosity $L$.
The almost all energy injected in the plasmas
escape as photon energy.
Outside the boundary, most of these photons may escape without
interaction with the pair-fluid.
Therefore, the pair-fluid does not behave as a fireball
in these cases, which is out of our interest.

As the luminosity $L$ increases above $10^{45}$ erg ${\rm s^{-1}}$,
the pair density overwhelms the proton density.
Although the pair temperature $\theta_\pm$ continues to decrease
with growth of $L$, $\theta_\pm$ is always larger than $0.1$.
For $L \gtrsim 10^{46}$ erg ${\rm s^{-1}}$,
a considerable amount of energy escapes in the form of the pair-fluid.
In this high-luminosity case, the outflow is driven by
radiative force rather than their own pressure.

\begin{figure}[t]
\centering
\epsscale{1.0}
\plotone{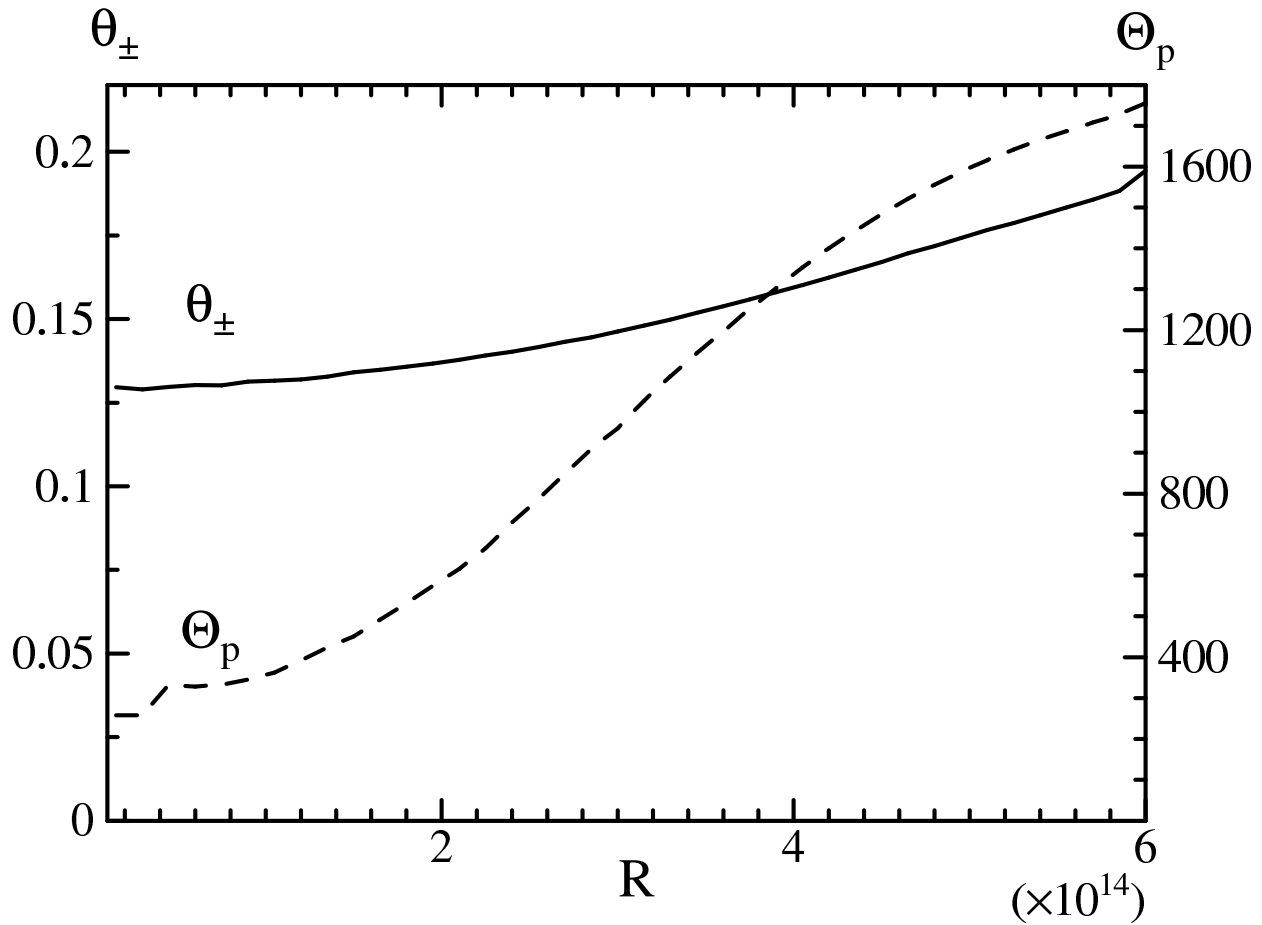}
\caption{
Temperatures $\theta_\pm$ (solid, left axis)
and $\Theta_{\it p} \equiv T_{\it p}/(m_{\it e} c^2)$ (dashed, right axis).
}
\end{figure}

In this paper we show detailed results for $L=10^{47}$ erg ${\rm s^{-1}}$,
$R_0=3 \times 10^{14}$ cm, and $n_0=\frac{1}{3} \times 10^{10}$
${\rm cm^{-3}}$ as a representative case.
The positron density $n_+$ and 4-velocity $U$ are plotted in Figure 1.
The ratio of number densities of positrons and protons $z \equiv \Gamma n_+/n_{\rm p}$
is 74 and 440 at the center and the boundary $R=R_{\rm out}$, respectively.
The 4-velocity $U$ monotonically increases with $R$,
and becomes 1.01 at the boundary.
The mean free path for the Thomson scattering at the boundary
is $l_{\rm T}=\Gamma/( n_\pm \sigma_{\rm T})=1.1 \times 10^{14}$ cm,
which is much shorter than $R=R_{\rm out}=6 \times 10^{14}$ cm.
Therefore, even outside the boundary,
the outflow and photons are tightly coupled and behave as a fireball.
In other words, our simulation is confined well inside the 
photosphere, which raises some problem with the treatment of the radiative 
transfer. However, as is seen later, 
since the photons are beamed in the outer region, the problem may not be 
so severe. 

The outflow may continue to be accelerated outside $R_{\rm out}$.
As conventionally used in the GRB standard model, we define
the ratio of the total luminosity to the mass ejection rate as
\begin{eqnarray}
\eta \equiv \frac{L}{4 \pi R_{\rm out}^2 n_\pm(R_{\rm out})
U(R_{\rm out}) m_{\rm e} c^3},
\end{eqnarray}
In the GRB models this value, which corresponds to
the final Lorentz factor of the outflow,
is defined using the baryon ejection rate.
In our case too, if the number of pairs is practically conserved
within the photospheric radius,
the final Lorentz factor of the flow is on the same order of $\eta$,
as demonstrated in \citet{iwa04}.
Our simulation shows $\eta \simeq 34$, which is enough to
explain the high Lorentz factor of AGN jets.
However, in order to conserve the number of pairs,
a relativistic photon temperature at the photosphere is required.

\begin{figure}[t]
\centering
\epsscale{1.0}
\plotone{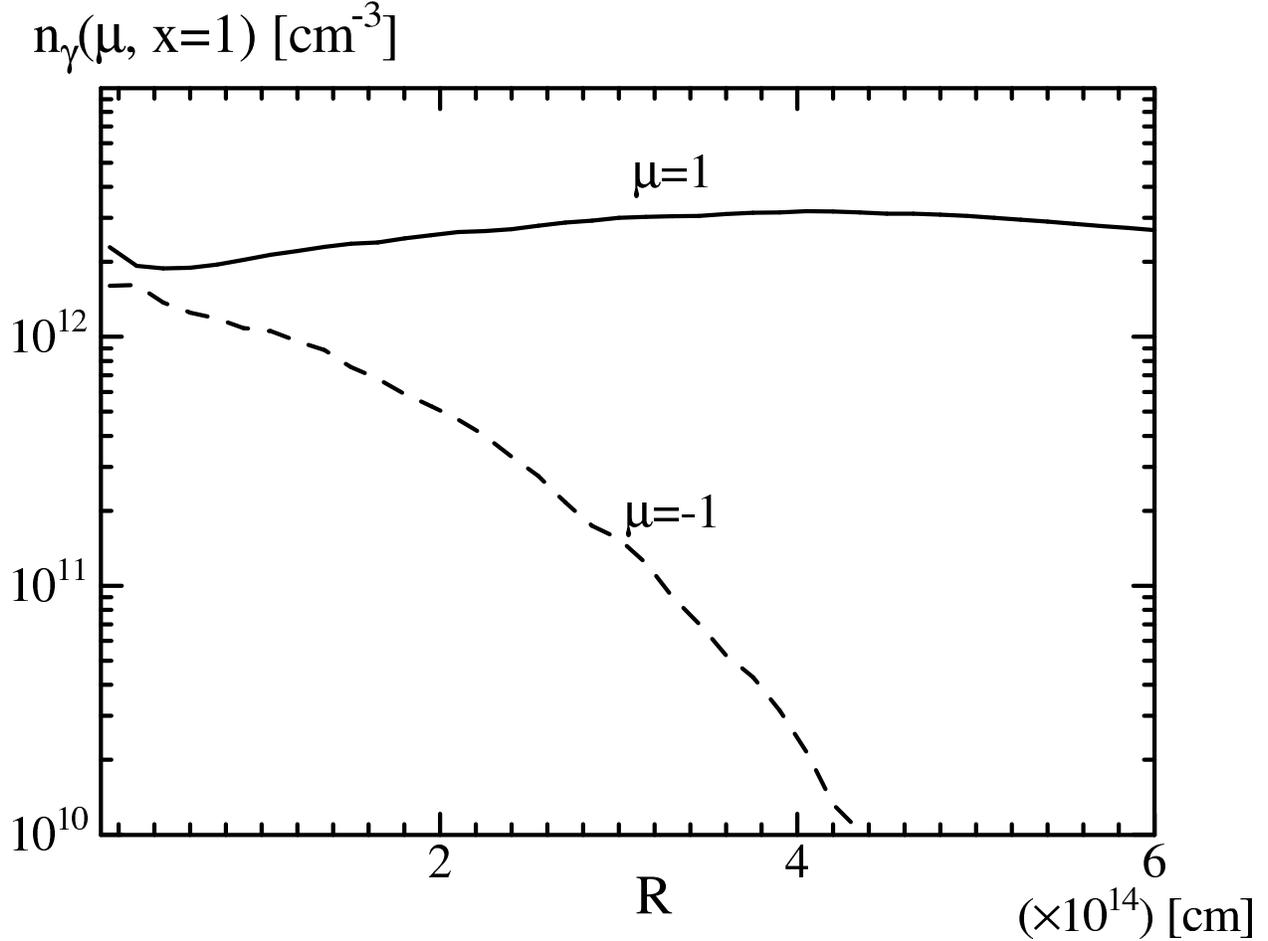}
\caption{
Photon densities of outgoing (solid) and ingoing (dashed)
photons for $x=1$ in the coordinate frame $K$.
}
\end{figure}

\begin{figure}[t]
\centering
\epsscale{1.0}
\plotone{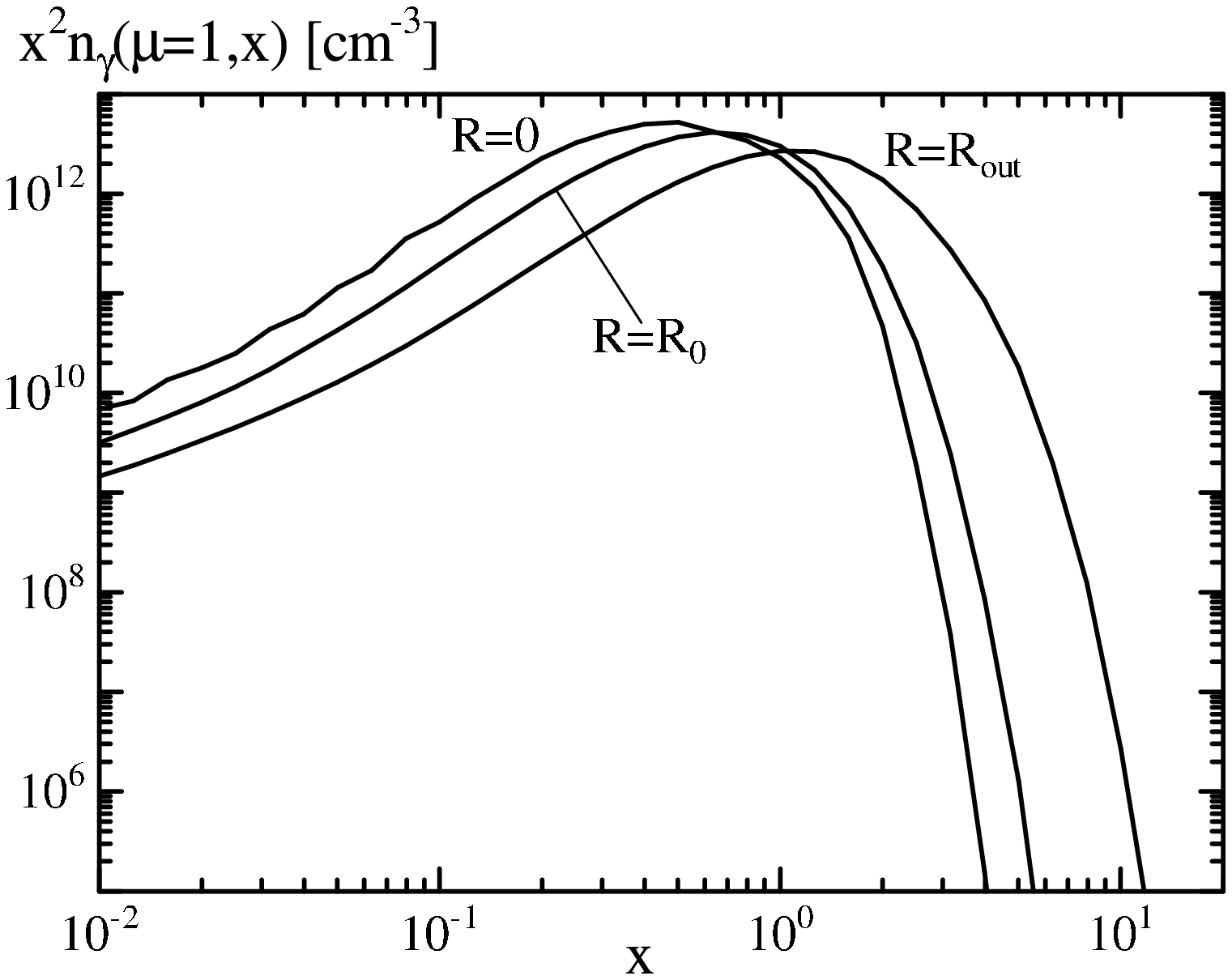}
\caption{
Photon spectra of outgoing photons for
$R=0$, $R_{\rm out}/2$, and $R_{\rm out}$ in the coordinate frame $K$.
}
\end{figure}

The temperatures $\theta_\pm$ and $\Theta_{\it p} \equiv T_{\it p}/(m_{\it e} c^2)$
are plotted in Figure 2.
At the center the pair temperature $\theta_\pm$ is 0.13.
The temperature $\theta_\pm$ increases with $R$ slightly,
and becomes 0.19 at the boundary,
which is smaller than the temperature assumed
in the Wien fireball model \citep{iwa02,iwa04}.
In this case, because of pair-annihilation,
the number of pairs may decrease seriously before pairs and photons decouple.
The temperature $\theta_{\it e}$ is almost the same as $\theta_\pm$
so that we do not plot $\theta_{\it e}$.
At the same time, the pair temperature does not show a 
adiabatic cooling characteristic of the fireball. 
This is due to a distributed Coulomb heating by protons.
If this heating is effective up to the photosphere, 
pair annihilation problem may be significantly alleviated.
In this example the proton temperature $\Theta_{\it p}$ at the center
is 260, which corresponds to $1.5 \times 10^{12}$ K.
On the other hand, at the boundary,
$\Theta_{\it p}$ becomes 1760 ($10^{13}$ K).
Since this proton temperature is too high to confine 
by gravitational potential of the central black hole, protons will also 
outflow from the surface, which is in contradiction with our 
basic assumption of a static proton distribution.

The outflow is due to the anisotropy of the photon field.
In Figure 3 we plot densities of outgoing ($\mu=1$)
and ingoing ($\mu=-1$) photons $n_\gamma (\mu,x) d \mu dx$
for $x=1$ in the frame $K$.
In the neighborhood of the center the densities
of outgoing and ingoing photons are almost the same
within errors in our computation.
As it goes outside, outgoing photons dominate over 
ingoing photons.
This is due to the spherical symmetry in geometry
and the beaming effect by the mildly relativistic outflow (see also Figure 4).
The very low density of ingoing photons at the boundary
assures the validity of our treatment in computation,
in which we neglect photons coming from outside of the boundary.
Because of the beaming effect, the density of outgoing photons
is almost constant in the frame $K$,
although the photon density in the comoving frame $K'$
decreases with $R$.
Figure 4 shows the photon spectra of outgoing photons
for $R=0$, $R_0=R_{\rm out}/2$, and $R_{\rm out}$.
The beaming effect is clearly seen in these plots.

\begin{figure}[t]
\centering
\epsscale{1.0}
\plotone{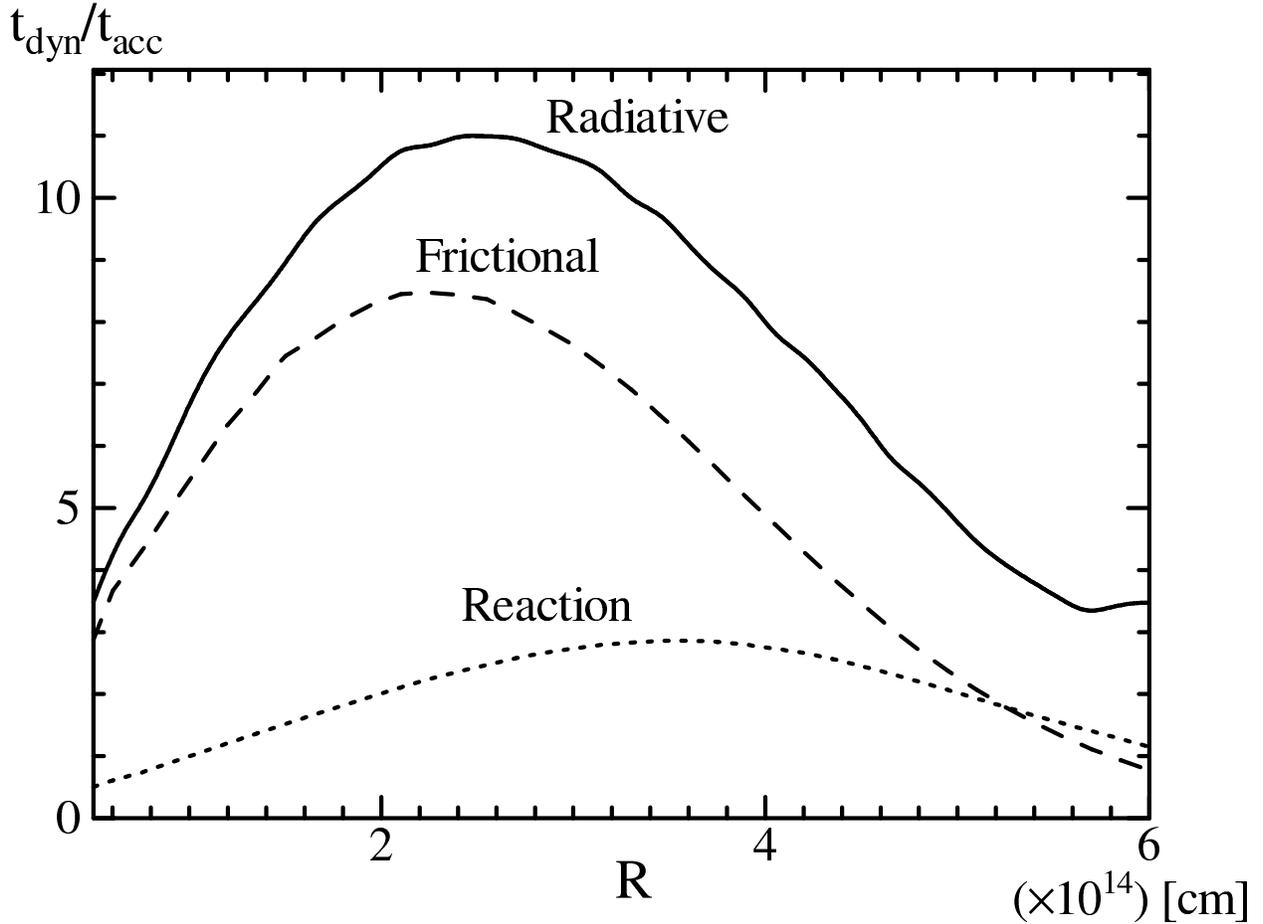}
\caption{
Radiative (solid), frictional (dashed) and reaction (dotted)
forces per particle.
}
\end{figure}

In Figure 5 we plot the radiative force, frictional force,
and reaction force due to photon-emission per particle
as ratios of timescales; the dynamical timescale $t_{\rm dyn} \equiv R_0/c$
to the acceleration (deceleration) timescale
$t_{\rm acc} \equiv \Gamma n_\pm m_{\rm e} c/{\cal F}$,
where ${\cal F}=F_\gamma$, $F_{\it e}+F_{\it p}$,
$F_{\rm rea}$ for the radiative, frictional,
and reaction forces, respectively.
The reaction force causes the reduction in the inertia of the flow
rather than flow-deceleration.
The radiative force also causes both the increase in the inertia
and flow-acceleration.
Thus, Figure 5 does not directly explain the growth of $U$,
but we can see a very powerful role of the radiative force.
Around the center, the radiative force per particle
increases with $R$ because of the anisotropy of the photon field.
At $R \sim R_0$, $t_{\rm dyn}/t_{\rm acc}$ due to the radiative force
attains the maximum $\sim 11$.
Outside $R_0$, the decline of the photon field in the frame $K'$
leads to the decline of the radiative force per particle.
In spite of the powerful radiative force,
$U$ does not attain the value
expected simply from $t_{\rm dyn}/t_{\rm acc} \sim 10$,
because the frictional force competes against the radiative force.
This strong friction will in turn accelerate the background 
proton-electron plasma. Although we assume that 
the background plasma is static, in reality it will outflow at 
a sub-relativistic speed. 
Although $U$ increases monotonically with $R$,
the frictional force decreases outside $\sim R_0$
owing to the smaller background density and
the cross sections of Coulomb scattering above $U \sim 1$.

\begin{figure}[t]
\centering
\epsscale{1.0}
\plotone{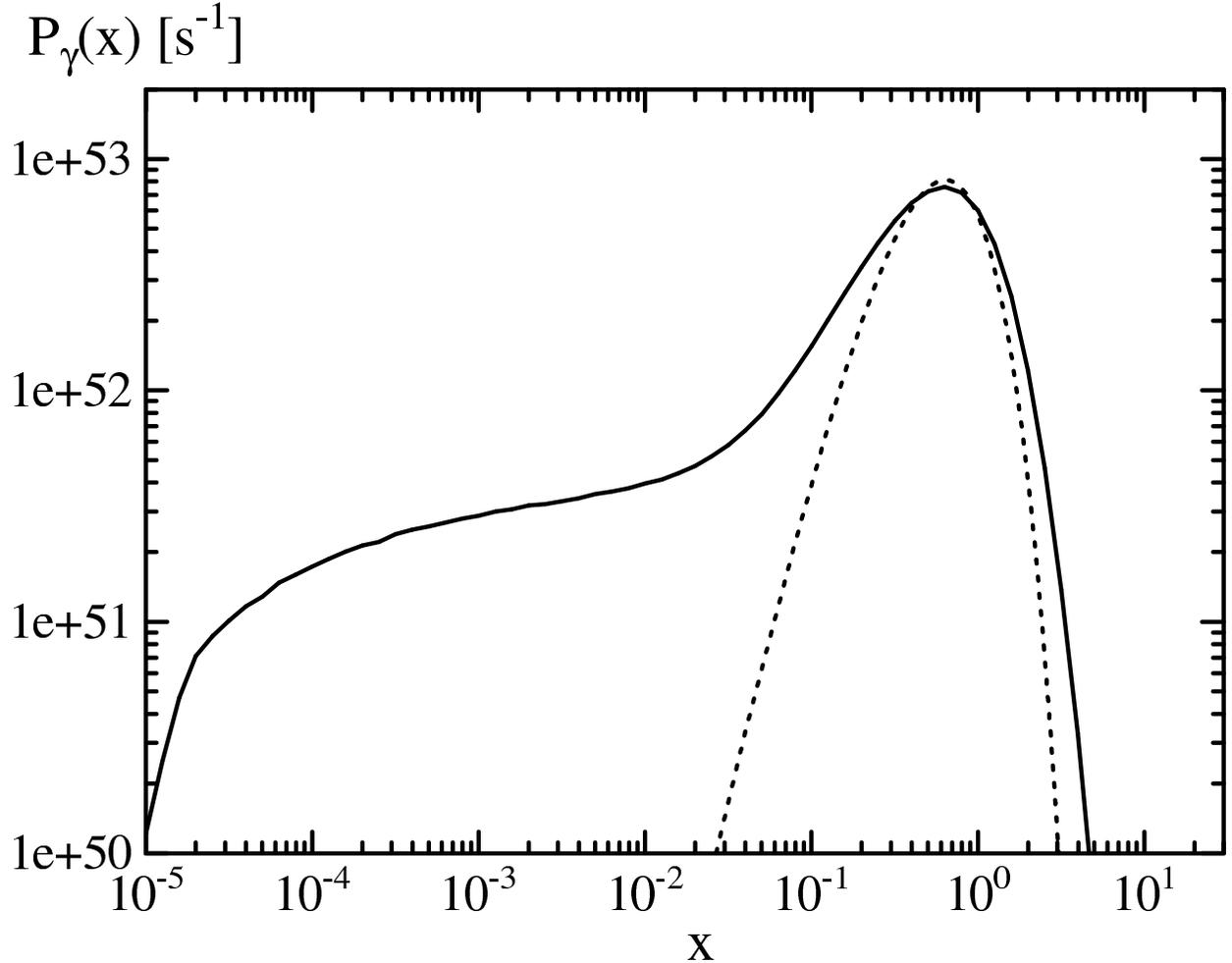}
\caption{
Energy spectrum of the photons escaping from the boundary (solid).
The dotted line is the Wien spectrum for reference.
}
\end{figure}

The energy outflow rate in the form of the pair-fluid
is $7.7 \times 10^{45}$ erg ${\rm s^{-1}}$ in this example.
The rest of the energy escapes as photons from the boundary.
In Figure 6 we plot the energy spectrum of the photons
escaping from the boundary $P_\gamma(x) dx$.
By the analogy from the Planck spectrum
we can obtain the ``photon temperature''
from the blue-shifted maximum point at $x=x_{\rm max}=0.63$
in the energy spectrum as
$\theta_\gamma=x_{\rm max}/(2.822 \Gamma)=0.15$,
which is close to the pair temperature $\theta_\pm$.
Therefore, we could not expect a rise of the pair temperature
outside the boundary.
For reference, we also plot the Wien spectrum
$\propto x^3 \exp{[-x/(\Gamma \theta_\gamma)]} dx$ in Figure 6.
The spectrum has a broader feature than that in the Wien spectrum.
Although we can see a sign of upscattering of soft photons
originating from bremsstrahlungs,
there are huge amounts of soft photons ($x<10^{-2}$)
in comparison with the thermal spectrum.

\begin{figure}[t]
\centering
\epsscale{1.0}
\plotone{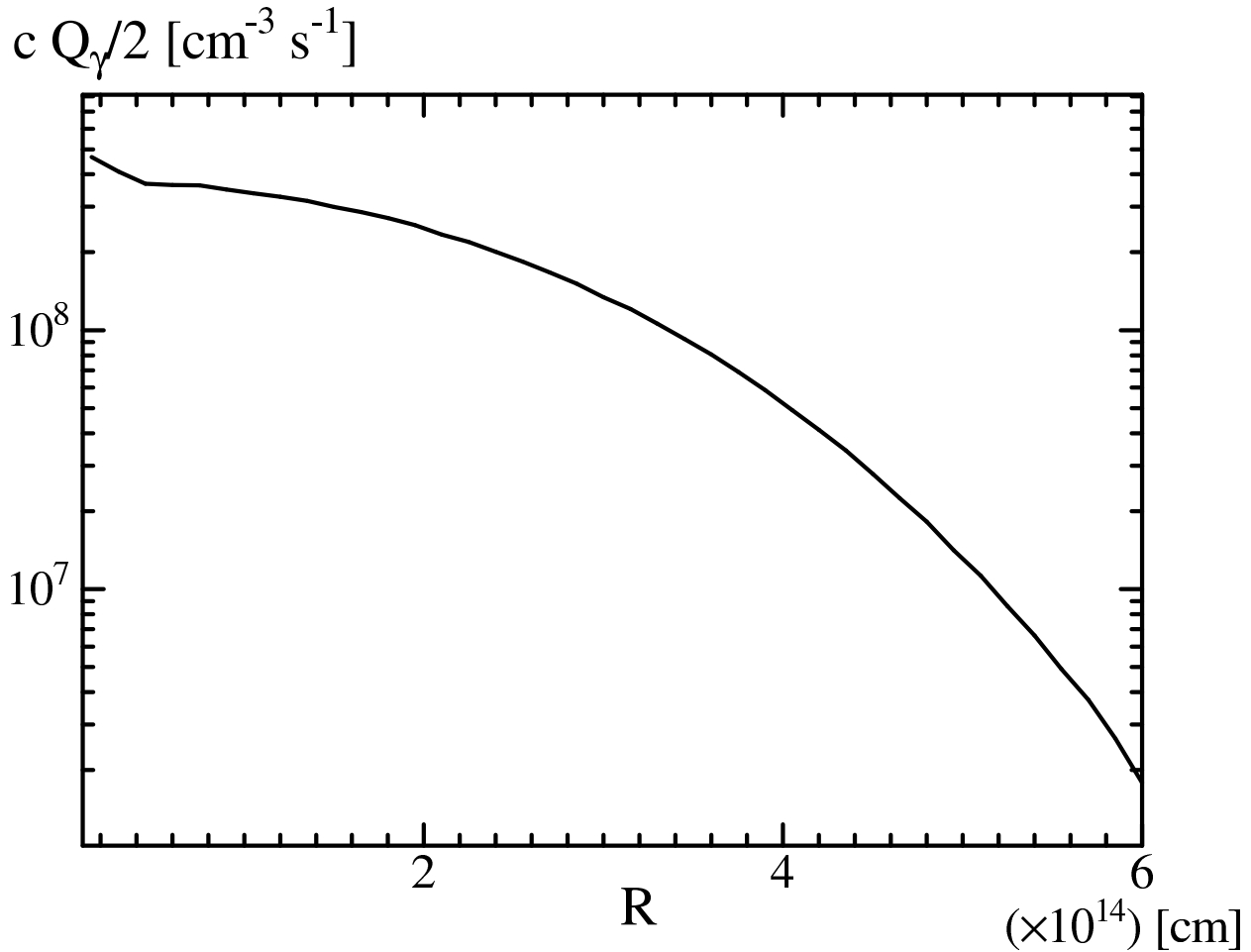}
\caption{
Positron creation rate.
}
\end{figure}

Despite of the low photon temperature,
the positron creation rate (see Figure 7)
and annihilation rate are almost the same within $R_{\rm out}$.
Therefore, we do not plot the positron annihilation rate
in the figure.
Outside the boundary the flow continues to be accelerated by the radiative force
until the photospheric radius.
Although the pair annihilation process exceeds the creation process
outside $R_{\rm out}$,
the number of pairs will be frozen at the photospheric radius.
Using the results in \citet{iwa02},
the Lorentz factor at the photosphere
may be $\tau_{\rm out}^{1/3} \sim 2$, where $\tau_{\rm out}$
is the Thomson optical depth at the boundary.
The final energy-flux fraction of the outflow to $L$
may be on the order of $\sim 10$ \%.

For $L=10^{47}$ erg ${\rm s^{-1}}$,
we have also simulated for another parameter set:
$R_0=10^{14}$ cm with $n_0=10^{10}$ ${\rm cm^{-3}}$.
However, the values $\theta_\pm$, $U$, $\eta$, and $\theta_\gamma$
at the boundary are almost the same as those in this example.
Therefore, the results may not be so sensitive to
the compactness of the plasma.

\section{CONCLUSIONS AND DISCUSSION}

This is the first attempt to produce fireballs from
hot plasmas, whose size and density are characteristic
of accretion plasma in AGNs.
For microphysics, Coulomb scattering, Compton scattering,
bremsstrahlung, electron-positron
pair annihilation and creation are taken into account.
Using the multi-fluid approximation, we obtain spherically symmetric,
steady solutions of radiation and pair outflows
for $L \leq 10^{47}$ erg ${\rm s^{-1}}$.
As the luminosity increases, sufficient amount of
electron-positron pairs outflow with a mildly relativistic velocity.
The pair outflows are driven by powerful radiative force rather than
their own pressure for $L > 10^{45}$ erg ${\rm s^{-1}}$.
Although we had expected that the dynamical effect of the outflow
would increase the temperature of the pair plasma,
in our results of $L > 10^{45}$ erg ${\rm s^{-1}}$,
the temperature is rather low to achieve $\Gamma \sim 10$,
differently from the assumption
in the Wien fireball model \citep{iwa02,iwa04}.
However, this difference should not taken too firmly 
at the present stage as is discussed below. 
 
There are several caveats in our present simulations. 
First, our calculation regime may not be large enough to 
accurately treat the whole spatial range within the photosphere. 
Distributed heating by protons in the outer region may affect the resultant 
pair outflow; one signature seen in the simulation is the lack of 
an adiabatic cooling in the flow. 
Secondly, the Coulomb friction rate and the resultant proton 
temperature suggest that protons will also outflow with sub-relativistic 
velocity. The flow of the background plasma will somewhat 
enhance the outflow of the pair-fluids.
If we can raise the temperature to only several times the value
in our simulation, the fireball can form a relativistic outflow.
The temperature, density, and velocity of the pair fluid
may be strongly influenced by spatial distributions of
the heating rate and proton density.
Qualitatively speaking, shallower proton density distribution 
will enhance the outflow, while more centrally concentrated 
heating rate will enhance the photon luminosity.
In order to produce a fireball with a sufficiently high temperature,
we need to continue investigating plasmas with different profiles 
including the flow of the background plasma.
While our simulation is one dimensional,
two or three dimensional effects may be important.

Another method we should consider
is kinematical treatment of the flow.
The multi-fluid approximation has been adopted in our simulation
for simplicity.
To carry out more accurate simulation,
we should solve collisional Boltzmann equation
taking into account photon emission.
The kinematical treatment may result in a broader distribution of pairs
than Maxwell-Boltzmann distribution,
which raises the effective temperature.
The broad photon distribution in our simulation suggests such a possibility.
The effect of the static electric field discussed in \S 2.3,
which may heat the pair-fluid,
can be included in the kinematical treatment.
However, such simulations require enormous cost in computation.
In addition, photon emission from plasmas with multi-component,
which is not in Maxwell-Boltzmann distribution,
is too complex to evaluate in advance of simulation.
Although the kinematical methods can include several plasma
effects we have neglected, such as two-stream instability etc.,
simulations based on the multi-fluid approximation
are irreplaceable way so far.

In this paper we have obtained steady solutions for outflows. 
Since the time scale of radiative transfer is much longer than the 
dynamical time scale, the pair and electron temperatures 
become low. The lifetime of accretion plasma may 
not be long enough to achieve a steady state radiation field 
obtained here. 
As the original idea of the Wien fireball model, 
an instantaneous heating may produce
a runaway production of relativistic pairs
because of slow annihilation in a hot plasma.
Therefore, time-dependent simulations should be
examined in a future work.

\begin{acknowledgments}
The authors appreciate the anonymous referee and
M. Kusunose for their helpful advice.
This work is partially supported by Scientific Research Grants
(F.T. 14079205 and 16540215) from the Ministry of Education, Culture,
Science and Technology of Japan. 
\end{acknowledgments}

\onecolumn

\end{document}